\documentclass[11pt,english]{article}
\usepackage[latin9]{inputenc}
\usepackage{geometry}
\usepackage{float}
\usepackage{graphicx}

\makeatletter
\newcommand{\lyxaddress}[1]{
	\par {\raggedright #1
	\vspace{1.4em}
	\noindent\par}
}

\makeatother

\usepackage{babel}
\begin{document}

\title{Effect of transcription reinitiation in stochastic gene expression}

\author{Rajesh Karmakar$^{*}$ and Amit Kumar Das$^{\#}$}
\maketitle

\lyxaddress{\begin{center}
$^{*}$Department of Physics, Ramakrishna Mission Vidyamandira, Belur
Math, Howrah-711202, India.$\qquad\qquad$ $\#$Kharial High School,
Kanaipur, Hooghly-712234, India.$\qquad\qquad\qquad\qquad\qquad\qquad\qquad\qquad\qquad\qquad\qquad$
\par\end{center}}
\begin{abstract}
Gene expression (GE) is an inherently random or stochastic or noisy
process. The randomness in different steps of GE, e.g., transcription,
translation, degradation, etc., leading to cell-to-cell variations
in mRNA and protein levels. This variation appears in organisms ranging
from microbes to metazoans. Stochastic gene expression has important
consequences for cellular function. The random fluctuations in protein
levels produce variability in cellular behavior. It is beneficial
in some contexts and harmful to others. These situations include stress
response, metabolism, development, cell cycle, circadian rhythms,
and aging. Different model studies e.g., constitutive, two-state,
etc., reveal that the fluctuations in mRNA and protein levels arise
from different steps of gene expression among which the steps in transcription
have the maximum effect. The pulsatile mRNA production through RNAP-II
based reinitiation of transcription is an important part of gene transcription.
Though, the effect of that process on mRNA and protein levels is very
little known. The addition of any biochemical step in the constitutive
or two-state process generally decreases the mean and increases the
Fano factor. In this study, we have shown that the RNAP-II based reinitiation
process in gene transcription can have different effects on both mean
and Fano factor at mRNA levels in different model systems. It decreases
the mean and Fano factor both at the mRNA levels in the constitutive
network whereas in other networks it can simultaneously increase or
decrease both quantities or it can have mixed-effect at mRNA levels.
We propose that a constitutive network with reinitiation behaves like
a product independent negative feedback circuit whereas other networks
behave as either product independent positive or negative or mixed
feedback circuit.

\end{abstract}

\section{Introduction}

Gene expression is a fundamental cellular process consisting of several
consecutive random steps like transcription, translation, degradation,
etc. The random nature of the biochemical steps of gene expression
is responsible for the stochastic or noisy production of mRNA and
protein molecules. This stochasticity in gene expression gives rise
to heterogeneity in an identical cell population and phenotypic variation.
Phenotypic variation is generally attributed to genetic and environmental
variation. Though it has been observed that genetically identical
cells in a constant environment show significant phenotypic variation. 

The origin and consequences of noise in stochastic gene expression
have been studied extensively, both theoretically and experimentally,
during the last three decades \cite{key-11,key-12,key-13,key-14,key-15,key-16,key-17,key-18,key-19,key-20,key-21,key-22,key-23,key-24,key-25,key-26,key-27,key-28,key-29,key-30}.
Several studies on both prokaryotic \cite{key-16,key-22} and eukaryotic
systems \cite{key-17,key-18,key-23,key-27} suggest that gene transcription
occurs in a discontinuous manner and that gives rise to fluctuating
production of mRNAs and proteins. The random fluctuations in the number
of mRNA and protein molecules in each cell constitute the noise. The
cells must either exploit it, learn to cope with it, or overcome it
using its internal noise suppression mechanisms. It can improve fitness
by generating cellular heterogeneity in clonal cell populations, thus
enabling a fast response to varying environments \cite{key-18}. Because
of its functional importance in cellular processes, it is necessary
and important to identify and dissect the biochemical processes that
generate and control the noise.

The transcription is an important step in stochastic gene expression.
It has been observed that the transcription process contributes maximum
noise in protein level than any other biochemical steps in gene expression
\cite{key-14,key-17,key-18,key-19,key-21,key-22,key-23,key-26,key-27}.
During the transcription process, different transcription factors
(TFs) bind to multiple sites on regulatory DNA in response to intracellular
or extracellular signals. On binding the regulatory systems, the TFs
turn the gene into an active state from which a burst of mRNAs is
produced. Transcriptional bursting has been observed across species
and is one of the primary causes of variability in gene expression
in cells and tissues \cite{key-17,key-18,key-20,key-22,key-23,key-31,key-32,key-33}.
Many experimental observations are modeled with that burst mechanism
\cite{key-19,key-21,key-24,key-26,key-28,key-34}. The mRNA synthesis
from the active gene actually takes place through interactions with
RNAP-II \cite{key-35,key-36,key-37,key-38}. Experiments show that
the RNAP-II based transcription, specific to eukaryotes, produces
pulsatile mRNA production through reinitiation and is crucial to reproduce
the experimental observations on noise at protein levels \cite{key-17,key-18}.
The Reinitiation of transcription introduces the third state of a
gene along with the two states of the two-state model network. Recent
research shows that the c-Fos gene in response to serum stimulation
indicates that a third state along with the inactive and active states
is essential to explain the experimental data on variance \cite{key-31}.
Noise strength or Fano factor at the mRNA level is unity for constitutive
gene expression. The two-state model shows the super-Poissonian Fano
factor at the mRNA level because of the random nature of the gene
activation and deactivation. Only negative feedback in the two-state
model network can reduce the noise in the mRNA levels but not below
the unity or sub-Poissonian level. Saho and Zeitlinger propose, based
on their experimental observations, that paused RNAP II prevents new
initiation of transcription which may reduce noise \cite{key-38A}.
Recent work on the two-state system with RNAP-II based transcriptional
reinitiation process shows that the reinitiation step in transcription
has the ability to reduce the noise strength or Fano factor at mRNA
level below unity \cite{key-39}. Reduction of noise strength can
have important biological significance as it has different functional
roles in cellular activity, development, and evolution \cite{key-21,key-25,key-39A}.
The average level of proteins and the fluctuations about the average
level can play crucial roles in different diseases \cite{key-39B,key-39C,key-39D,key-39E}.
An appropriate amount of both, the mean and noise strength at mRNA
and protein levels, are therefore important for fine-tuned robust
cellular processes \cite{key-39F}. 

In this paper, we have considered different gene expression models
e.g., constitutive \cite{key-30}, two-state \cite{key-21}, and Suter
\cite{key-27} models, and studied the effect of transcription reinitiation
on the mean and Fano factor at mRNA levels. The stochastic events
at the gene activation and transcription process generally decrease
the average and increase the Fano factor at mRNA and protein levels.
But we show that reinitiation of transcription has the ability to
control the average and Fano factor both at mRNA and protein levels
in different combinations viz. either it decreases the both or increases
the both or shows mixed-effect (mean increases and Fano factor decreases
or vice-versa) in different model systems. From our exact analytical
calculations, we find that the reinitiation of transcription behaves
like product-independent negative feedback in the constitutive gene
regulatory network. Whereas for the two-state and Suter model, the
RNAP-II based transcription reinitiation behaves as either positive
or negative or mixed feedback circuit depending on the rate constants
of the biochemical steps. That is, the reinitiation process in gene
transcription can simultaneously control the mean and Fano factor
both at mRNA level. 

\section{Different gene expression models and analysis}

\subsection{Constitutive gene expression}

\begin{figure}[H]
\begin{centering}
\includegraphics[width=5cm,height=2.2cm]{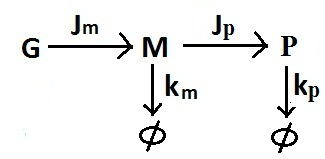}
\par\end{centering}
\caption{\label{fig:Constitutive}Reaction scheme with rate constants for constitutive
gene expression. $J_{m}$ ($J_{p}$) is the transcription (translation)
rate constant and $k_{m}$ ($k_{p}$) is the mRNA (protein) degradation
rate constant. }
\end{figure}

The essential genes in the cell always produce mRNAs and proteins.
The expression from essential genes is modeled by the constitutive
network shown in figure (\ref{fig:Constitutive}). In that model,
the gene is always assumed to be at the active state from which the
mRNA synthesis takes place with rate constant $J_{m}$. The proteins
are then synthesized from the newly born mRNAs. Both the mRNAs and
proteins are degraded with rate constants $k_{m}$ and $k_{p}$ respectively.
It is very easy to find out the expressions of mean, variance, and
Fano factor for the constitutive gene expression at the steady-state
by the Master equation approach \cite{key-40}.

Let $p(n_{1},n_{2},t)$ be the probability density of $n_{1}$ mRNAs
and $n_{2}$ proteins at time $t$. The rate of change of probability
is given by the Master equation \cite{key-40}
\begin{equation}
\begin{array}{ccc}
\frac{\partial p(n_{1},n_{2},t)}{\partial t} & = & l\,J_{m}[p(n_{1}-1,n_{2},t)-p(n_{1},n_{2},t)]\\
 &  & +k_{m}[(n_{1}+1)p(n_{1}+1,n_{2},t)-n_{1}p(n_{1},n_{2},t)]\\
 &  & +J_{p}[n_{1}p(n_{1},n_{2}-1,t)-n_{1}p(n_{1},n_{2},t)]\\
 &  & +k_{p}[(n_{2}+1)p(n_{1},n_{2}+1,t)-n_{2}p(n_{1},n_{2},t)]
\end{array}\label{eq:C-1}
\end{equation}

where $l$ is the copy number of the gene.

The steady state solution of equation (\ref{eq:C-1}) for the constitutive
gene expression process gives the mean ($<m^{c}>$) , variance ($Var_{m}^{c}$,
$Var_{p}^{c}$) and Fano factor ($FF_{m}^{c}$) of mRNAs and Fano
factor ($FF_{p}^{c}$) of proteins and are given by (for $l=1$)
\begin{center}
\begin{equation}
<m^{c}>=\frac{J_{m}}{k_{m}};\quad<p^{c}>=<m^{c}>\frac{J_{p}}{k_{p}}\label{eq:C-2}
\end{equation}
\par\end{center}

\begin{equation}
Var_{m}^{c}=\frac{J_{m}}{k_{m}},\quad FF_{m}^{c}=\frac{Var_{m}^{c}}{<m^{c}>}=1\label{eq:C-3}
\end{equation}

\begin{equation}
Var_{p}^{c}=<p^{c}>\frac{J_{p}+k_{m}+k_{p}}{k_{m}+k_{p}},\quad FF_{p}^{c}=\frac{Var_{p}^{c}}{<p^{c}>}=\frac{J_{p}+k_{m}+k_{p}}{k_{m}+k_{p}}\label{eq:C-4}
\end{equation}

The noise strength or Fano factor of mRNAs in constitutive GE is unity.
That is a unique feature of the Poisson process and that can be taken
as a reference to compare with other gene expression network models. 

\subsection{Constitutive gene expression with reinitiation}

\begin{figure}[H]
\begin{centering}
\includegraphics[width=5cm,height=2.2cm]{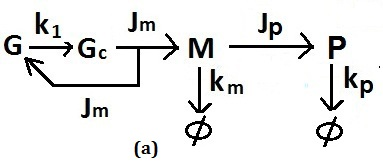}
\includegraphics[width=5cm,height=2.2cm]{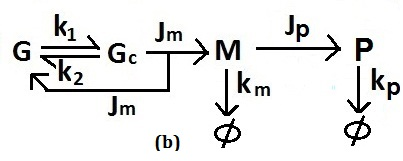}
\par\end{centering}
\caption{\label{fig:Constitutive with reinitiation}Reaction schemes with rate
constants for constitutive gene expression with reinitiation (a) without
reverse reaction and (b) with reverse reaction. The RNAP-II binds
the gene (G) with rate constant $k_{1}$ and forms an initiation complex
($G_{c}$). $k_{2}$ is the dissociation rate constant of RNAP-II
from the initiation complex. $J_{m}$ ($J_{p}$) is the transcription
(translation) rate constant and $k_{m}$ ($k_{p}$) is the mRNA (protein)
degradation rate constant. }
\end{figure}

In constitutive gene expression, the binding and movement of RNAP-II
are ignored. But in the actual process, the RNAP-II molecules bind
the gene to form an initiation complex \cite{key-35}. In the next
step, the bound RNAP-II leaves the initiation complex and starts transcription
along the gene. The gene then comes again into its normal state (Fig.
\ref{fig:Constitutive with reinitiation}(a)). In that process, it
is assumed that bound RNAP-II must do transcription without any uncertainty.
Though, that may not be possible always. There must be a finite probability
that bound RNAP-II leaves the initiation complex without transcribing
the gene. That is considered in figure \ref{fig:Constitutive with reinitiation}(b). 

To calculate the mean and variance/Fano factor, we consider the most
general reaction scheme in figure \ref{fig:Constitutive with reinitiation}(b).
Let $p(n_{1},n_{2},n_{3},t)$ be the probability density of $n_{1}$
genes in the $G_{c}$ state, $n_{2}$ mRNAs and $n_{3}$ proteins
at time $t$. The rate of change of probability density corresponding
to the reaction scheme in figure 2(b) is given by the Master equation
\cite{key-40}
\begin{equation}
\begin{array}{ccc}
\frac{\partial p(n_{1},n_{2},n_{3},t)}{\partial t} & = & k_{1}[\{l-(n_{1}-1)\}p(n_{1}-1,n_{2},n_{3},t)-(l-n_{1})p(n_{1},n_{2},n_{3},t)]\\
 &  & +k_{2}[(n_{1}+1)p(n_{1}+1,n_{2},n_{3},t)-n_{1}p(n_{1},n_{2},n_{3},t)]\\
 &  & +J_{m}[(n_{1}+1)p(n_{1}+1,n_{2}-1,n_{3},t)-n_{1}p(n_{1},n_{2},n_{3},t)]\\
 &  & +k_{m}[(n_{2}+1)p(n_{1},n_{2}+1,n_{3},t)-n_{2}p(n_{1},n_{2},n_{3},t)]\\
 &  & +J_{p}[n_{2}p(n_{1},n_{2},n_{3}-1,t)-n_{2}p(n_{1},n_{2},n_{3},t)]\\
 &  & +k_{p}[(n_{3}+1)p(n_{1},n_{2},n_{3}+1,t)-n_{3}p(n_{1},n_{2},n_{3},t)]
\end{array}\label{eq:C-5}
\end{equation}

Equation (\ref{eq:C-5}) gives the mean, variance, and Fano factor
at the steady state as (for $l=1$)

\begin{equation}
<m_{r}^{cwr}>=\frac{k_{1}}{J_{m}+k_{1}+k_{2}}\frac{J_{m}}{k_{m}};\quad<p_{r}^{cwr}>=<m_{r}^{cwr}>\frac{J_{p}}{k_{p}}\label{eq:C-6}
\end{equation}

\begin{equation}
Var_{m_{r}}^{cwr}=\frac{J_{m}}{k_{m}}\frac{k_{1}\{(J_{m}+k_{1}+k_{2})(J_{m}+k_{m}+k_{2})+k_{1}^{2}+k_{1}k_{2}\}}{(J_{m}+k_{1}+k_{2})^{2}(J_{m}+k_{1}+k_{2}+k_{m})}=<m_{r}^{cwr}>(1-\frac{J_{m}k_{1}}{(J_{m}+k_{1}+k_{2})(J_{m}+k_{1}+k_{2}+k_{m})})\label{eq:C-7}
\end{equation}

\begin{equation}
FF_{m_{r}}^{cwr}=1-\frac{J_{m}k_{1}}{(J_{m}+k_{1}+k_{2})(J_{m}+k_{1}+k_{2}+k_{m})}\label{eq:C-8}
\end{equation}

\begin{equation}
FF_{p_{r}}^{cwr}=\frac{Var_{p}^{cwr}}{<p^{cwr}>}=1+\frac{J_{p}\,g_{1}}{(J_{m}+g_{2})(J_{m}+g_{2}+k_{m})(J_{m}+g_{2}+k_{p})g_{3}}\label{eq:C-9}
\end{equation}

where $g_{1}=J_{m}^{3}+g_{2}(g_{2}+k_{m})(g_{2}+k_{p})+J_{m}^{2}(2k_{1}+3k_{2}+g_{3})+J_{m}\{2k_{1}^{2}+3k_{2}^{2}+k_{m}k_{p}+2k_{2}g_{3}+k_{1}(5k_{2}+g_{3})\}$,
$g_{2}=k_{1}+k_{2}$, $g_{3}=k_{m}+k_{p}$

\begin{figure}[H]
\begin{centering}
\includegraphics[width=5.5cm,height=3.5cm]{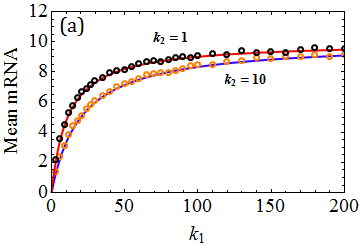}$\;$\includegraphics[width=5.5cm,height=3.5cm]{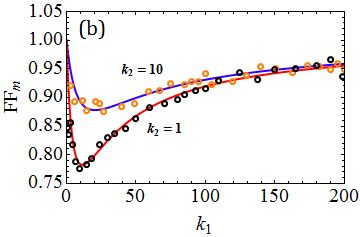}
\includegraphics[width=5.5cm,height=3.5cm]{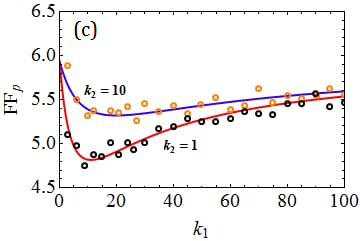}
\par\end{centering}
\caption{\label{fig:Variation-of-mean for Cons W Reini}Variation of (a) mean
mRNA, (b) Fano factor at mRNA levels and (c) Fano factor at protein
levels with $k_{1}$ for different values of $k_{2}$ with $J_{m}=10$
and $k_{m}=1$. The solid lines are drawn from analytical calculations
and hollow circles are generated from the stochastic simulation based
on the Gillespie algorithm \cite{key-40A}. }
\end{figure}

It is seen from equations (\ref{eq:C-6}), (\ref{eq:C-7}) and (\ref{eq:C-8})
that the transcriptional reinitiation in constitutive GE process (figure
\ref{fig:Constitutive with reinitiation}(a) and \ref{fig:Constitutive with reinitiation}(b))
decreases the mean and Fano factor at mRNA and protein levels in comparison
to that the constitutive GE process without reinitiation (Fig. \ref{fig:Constitutive}).
Figures \ref{fig:Variation-of-mean for Cons W Reini}(a), \ref{fig:Variation-of-mean for Cons W Reini}(b)
and \ref{fig:Variation-of-mean for Cons W Reini}(c) show that the
effect of reinitiation is strong enough at the lower values of $k_{1}$.
We see from the figures that the mean mRNA level and $FF_{m}$ approaches
the value observed in the constitutive process for a given value of
$J_{m}$ and $k_{m}$ for higher $k_{1}$. When $k_{1}$ increases
from zero value, the Fano factor at mRNA levels decreases and attains
a minimum value and then moves towards unity. The minimum of the Fano
factor ($FF_{m}$) will occur at $k_{1}=\sqrt{(J_{m}+k_{2})(J_{m}+k_{2}+k_{m})}$
(Fig. \ref{fig:Variation-of-mean for Cons W Reini}(b)). The equation
(\ref{eq:C-8}) shows that the Fano factor has identical dependence
on $k_{1}$ and $J_{m}$. As the rate constant $k_{2}$ increases,
the degree of deviation of the Fano factor below unity decreases (Fig.
\ref{fig:Variation-of-mean for Cons W Reini}(b)) because that decreases
the mean mRNA levels (Fig. \ref{fig:Variation-of-mean for Cons W Reini}(a)).
If one considers pre-initiation and initiation complexes in the transcriptional
reinitiation process rather than only the initiation complex then
the Fano factor further reduces below unity (Appendix A). 

To understand the effect of reinitiation of transcription on average
and Fano factor at mRNA level, we consider the constitutive gene expression
with reinitiation but without the reverse reaction ($k_{2}=0$) (Fig.
\ref{fig:Constitutive with reinitiation}(a)). We see that the reinitiation
in transcription reduces the mean mRNA level because the effective
transcription rate is reduced by a factor $\frac{k_{1}}{J_{m}+k_{1}}$.
That causes the reduction of variance at mRNA level. As a result,
the noise strength or Fano factor (Eq.(\ref{eq:C-8})) reduces below
unity with the reinitiation of transcription. 

The expression for mean mRNA (Eq. \ref{eq:C-6}) with $k_{2}=0$ can
be written as 

\begin{equation}
<m^{cwr}>=\frac{k_{1}}{J_{m}+k_{1}}\frac{J_{m}}{k_{m}}=\frac{<m^{c}>}{1+\beta<m^{c}>}\label{eq:C-10}
\end{equation}

This expression (Eq. \ref{eq:C-10}) is identical to the gain of a
linear negative feedback amplifier with the feedback factor $\beta=\frac{k_{m}}{k_{1}}$
\cite{key-40B}. The expression for the Fano factor (Eq. \ref{eq:C-8})
also shows that the noise strength is reduced with the reinitiation
of transcription. 

We can also write from the equation (\ref{eq:C-10})

\begin{equation}
\frac{d<m^{cwr}>}{<m^{cwr}>}=\frac{1}{1+\beta<m^{c}>}(\frac{d<m^{c}>}{<m^{c}>})\label{eq:C-11}
\end{equation}

Equation (\ref{eq:C-11}) shows that the percentage change in the
mean mRNA levels with reinitiation is much less than that without
reinitiation. That is reflected in the expression of the Fano factor
of mRNA with transcriptional reinitiation (Eq. (\ref{eq:C-8})). Therefore,
as far as the mean and Fano factor is concerned, the equations (\ref{eq:C-6})
and (\ref{eq:C-8}) (with $k_{2}=0$) and the equations (\ref{eq:C-10})
and (\ref{eq:C-11}) clearly indicate that the reinitiation of transcription
in gene expression behaves as a negative feedback loop in the regulatory
network. It is important to note that the negative feedback in the
gene regulatory network due to the reinitiation of transcription is
product independent. It is entirely inherent to the gene transcription
regulatory network because the mRNA and protein numbers do not have
any role in its synthesis here. The deterministic rate equation for
a gene regulatory network with an ordinary feedback loop, negative
or positive, contains the product-dependent synthesis part \cite{key-40C,key-40D}
but here the synthesis term is product independent and depends only
on the reaction rate constants.

The mean mRNA with reverse reaction (Eq.(\ref{eq:C-6})) can be expressed
as

\begin{equation}
<m_{r}^{cwr}>=\frac{<m^{c}>}{1+\beta<m^{c}>+\alpha<m^{c}>}\label{eq:C-12}
\end{equation}

where $\alpha=\frac{k_{m}}{k_{1}}\frac{k_{2}}{J_{m}}$. 

From equations (\ref{eq:C-6}), (\ref{eq:C-8}) and (\ref{eq:C-12}),
we see that the rate constant $k_{2}$ helps to reduce the mean mRNA
level further but increases the Fano factor (Figs. \ref{fig:Variation-of-mean for Cons W Reini}(b)
and (c)). So, the reverse transition with rate constant $k_{2}$ behaves
like negative feedback for mean mRNA level but positive feedback for
the Fano factor. Thus, the successful reinitiation of transcription
behaves like a negative feedback loop whereas the unsuccessful reinitiation
of transcription behaves like a mixed feedback loop. The effect of
reinitiation in gene expression is observed at the mRNA level first.
The nature of the variation in the Fano factor at the protein level
is the same as that at mRNA level except for a change in scale (Fig.
\ref{fig:Variation-of-mean for Cons W Reini}(c)). Therefore, the
Fano factor at the protein level does not give any new information
about the effect of reinitiation. So we keep our analysis up to the
mRNA level in the rest of the paper. 

\subsection{Two-state gene expression without and with transcriptional reinitiation
process}

\begin{figure}[H]
\begin{centering}
\includegraphics[width=5cm,height=2.2cm]{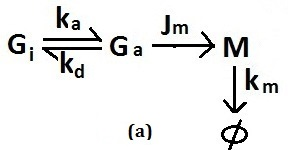}$\;\;$\includegraphics[width=6cm,height=2.2cm]{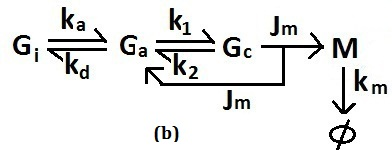}
\par\end{centering}
\caption{\label{fig:Two-state}Reaction scheme with rate constants for (a)
Two-state gene expression model and (b) Two-state gene expression
with reinitiation of transcription model. $k_{a}$ ($k_{d}$) is the
activation (deactivation) rate constant. $k_{1}$ is the rate constant
of initiation complex formation and $k_{2}$ is the rate constant
of dissociation of RNAP-II from initiation complex. $J_{m}$ is the
transcription rate constant and $k_{m}$ is the mRNA degradation rate
constant. }
\end{figure}

Regulation is ubiquitous in biological processes. The regulated gene
expression without feedback in a cellular system is modeled by the
two-state process. Many experimental results are explained with the
help of two-state gene expression process \cite{key-19,key-20,key-21,key-22,key-23,key-24,key-29}.
In that process, the gene can be in two possible states, active ($G_{a}$)
and inactive ($G_{i}$) (Fig. \ref{fig:Two-state}(a)) and random
transitions take place between the states. The mRNA synthesis occurs
in bursts only from the active state of the gene. The mRNAs have a
specific decay rate also. 

Now, let us assume that there is $l$ copy number of a particular
gene that exists in the cell. Let $p(n_{1},n_{2},t)$ be the probability
that at time $t$ and there are $n_{2}$ number of mRNAs with $n_{1}$
number of genes in the active state ($G_{a}$). The number of gene
in the inactive states are $(l-n_{1})$. The time evaluation of the
probability corresponding to the chemical reactions in figure \ref{fig:Two-state}(a)
is given by the Master equation \cite{key-40}

\begin{equation}
\begin{array}{ccc}
\frac{\partial p(n_{1},n_{2},t)}{\partial t} & = & k_{a}[(l-n_{1}+1)p(n_{1}-1,n_{2},t)-(l-n_{1})p(n_{1},n_{2},t)]\\
 &  & +k_{d}[(n_{1}+1)p(n_{1}+1,n_{2},t)-n_{1}p(n_{1},n_{2},t)]\\
 &  & +J_{m}[n_{1}p(n_{1},n_{2}-1,t)-n_{1}p(n_{1},n_{2},t)]\\
 &  & +k_{m}[(n_{2}+1)p(n_{1},n_{2}+1,t)-n_{2}p(n_{1},n_{2},t)]
\end{array}\label{eq:TS-1}
\end{equation}

Solving the equation (\ref{eq:TS-1}), we can easily find out the
mean, variance and Fano factor of mRNAs. They are given by

\begin{equation}
<m^{tswtr}>=\frac{k_{a}\:\;J_{m}}{(k_{a}+k_{d})k_{m}}\label{eq:TS-2}
\end{equation}

\begin{equation}
FF_{m}^{tswtr}=1+\frac{J_{m}k_{d}}{(k_{a}+k_{d})(k_{m}+k_{a}+k_{d})}\label{eq:TS-3}
\end{equation}

Equations (\ref{eq:TS-2}) and (\ref{eq:TS-3}) show that the inclusion
of inactive state and the random transitions between inactive and
active states in the constitutive process reduces the mean and increases
the Fano factor. 

In figure \ref{fig:Two-state}(b), we consider the reinitiation of
transcription along with the random transitions between the active
and inactive states of the gene. In the transcription reinitiation
step, an RNAP-II binds the gene in the active state and form an initiation
complex $G_{c}.$ Now, the bound RNAP-II has two choices, either it
moves forward or backward. If it moves forward then again two events
occur: mRNA synthesis and free up of the initiation complex. That
is, the initiation complex again becomes an active state where free
RNAP-II molecules can bind. The unsuccessful movement of RNAP-II from
the initiation complex of the gene brings it back to the active state
by dissociating the enzyme molecules. In the two-state gene expression
model, the randomness due to the transcriptional reinitiation process
is neglected assuming its insignificant contribution in mean and noise
strength at the mRNA and protein levels. Only a few works pointed
out that the reinitiation of transcription plays important role in
the phenotypic consequences of cellular systems \cite{key-17,key-18,key-39,key-41,key-41A}.
The above-mentioned gene expression model with reinitiation is studied
in ref \cite{key-39}. But for the completeness of this paper, we
are writing down here the Master equation and the expressions of means,
variances, and Fano factors. 

Let $p(n_{1},n_{2},n_{3},t)$ be the probability that at time $t$
and there are $n_{3}$ number of mRNAs with $n_{1}$ number of genes
in the active state ($G_{a}$) and $n_{2}$ number of genes in the
initiation state ($G_{c}$). The number of gene in the inactive states
are $(l-n_{1}-n_{2})$ with $l$ be the copy number of the gene. The
time evaluation of the probability corresponding to the biochemical
reactions in figure \ref{fig:Two-state}(b) is given by 

\begin{equation}
\begin{array}{ccc}
\frac{\partial p(n_{1},n_{2},n_{3},t)}{\partial t} & = & k_{a}[(l-n_{1}-n_{2}+1)p(n_{1}-1,n_{2},n_{3},t)-(l-n_{1}-n_{2})p(n_{1},n_{2},n_{3},t)]\\
 &  & +k_{d}[(n_{1}+1)p(n_{1}+1,n_{2},n_{3},t)-n_{1}p(n_{1},n_{2},n_{3},t)]\\
 &  & +k_{1}[(n_{1}+1)p(n_{1}+1,n_{2}-1,n_{3},t)-n_{1}p(n_{1},n_{2},n_{3},t)]\\
 &  & +k_{2}[(n_{2}+1)p(n_{1}-1,n_{2}+1,n_{3},t)-n_{2}p(n_{1},n_{2},n_{3},t)]\\
 &  & +J_{m}[(n_{2}+1)p(n_{1}-1,n_{2}+1,n_{3}-1,t)-n_{2}p(n_{1},n_{2},n_{3},t)]\\
 &  & +k_{m}[(n_{3}+1)p(n_{1},n_{2},n_{3}+1,t)-n_{3}p(n_{1},n_{2},n_{3},t)]
\end{array}\label{eq:TS-6}
\end{equation}

The expressions of averages and Fano factors of mRNAs for the reaction
scheme with the transcriptional reinitiation process in figure \ref{fig:Two-state}(b)
are given by \cite{key-39}

\begin{equation}
<m^{tswr}>=\frac{k_{a}k_{1}}{a_{2}}\frac{J_{m}}{k_{m}};\label{eq:TS-7}
\end{equation}

\begin{equation}
FF_{m}^{tswr}=1+\frac{J_{m}k_{1}(a_{2}-k_{a}a_{1})}{a_{2}(a_{1}k_{m}+a_{2})}\label{eq:TS-8}
\end{equation}

where $a_{1}=k_{m}+J_{m}+k_{a}+k_{d}+k_{1}+k_{2}$ and $a_{2}=k_{a}J_{m}+k_{d}J_{m}+k_{d}k_{2}+k_{1}k_{a}+k_{a}k_{2}$.

\begin{figure}[H]
\begin{centering}
\includegraphics[width=5.5cm,height=3.5cm]{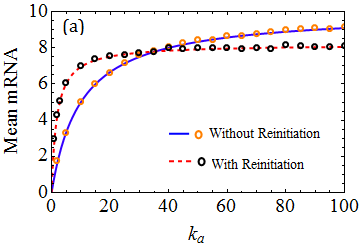}$\:$\includegraphics[width=5.5cm,height=3.5cm]{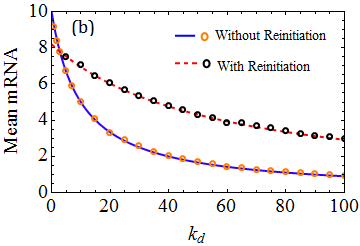}$\:$\includegraphics[width=5.5cm,height=3.5cm]{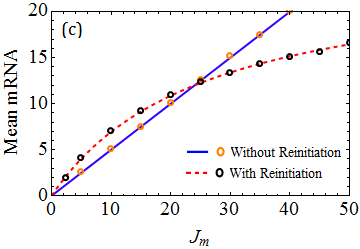}
\par\end{centering}
\caption{\label{fig:Variation-of-mean Two-state}Variation of mean mRNA levels
with and without reinitiation as a function of (a) $k_{a}$ , (b)
$k_{d}$ and (c) $J_{m}$ corresponding to the figure (\ref{fig:Two-state}).
The solid (dashed) lines are drawn from exact analytical expressions
in equation (\ref{eq:TS-2}) (Eq. (\ref{eq:TS-7})). The hollow circles
are generated using the stochastic simulation based on the Gillespie
algorithm. The rate constants are $k_{1}=50,$$k_{2}=1,$$k_{m}=1$,
$k_{d}=10$ and $J_{m}=10$ in figure (a) , $k_{a}=10$ and $J_{m}=10$,
in figure (b) and $k_{a}=10$ and $k_{d}=10$ in figure (c). }
\end{figure}

\begin{figure}[H]
\begin{centering}
\includegraphics[width=5.5cm,height=3.5cm]{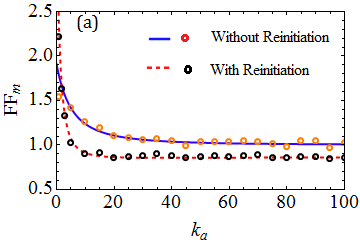} \includegraphics[width=5.5cm,height=3.5cm]{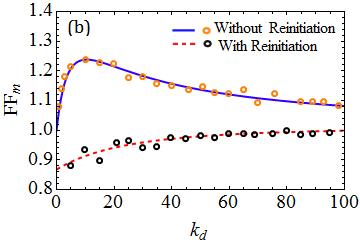}
\includegraphics[width=5.5cm,height=3.5cm]{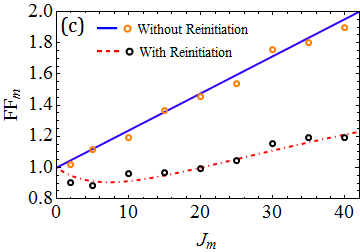}
\par\end{centering}
\caption{\label{fig:Plot-of-Fano-Two-state}Plot of Fano factor at mRNA level
versus (a) $k_{a}$ , (b) $k_{d}$ and (c) $J_{m}$ with and without
reinitiation. The solid (dashed) lines are drawn from exact analytical
expressions (Eqs. (\ref{eq:TS-3}) and (\ref{eq:TS-8})). The hollow
circles are obtained from stochastic simulation based on the Gillespie
algorithm. The rate constants are $k_{1}=50,$ $k_{2}=1,$ $k_{m}=1$,
$k_{d}=10$ and $J_{m}=10$ in figure (a) , $k_{a}=10$ and $J_{m}=10$
in figure (b) and $k_{a}=10$ and $k_{d}=10$ in figure (c). }

\end{figure}

\begin{figure}[H]
\begin{centering}
\includegraphics[width=5.5cm,height=3.5cm]{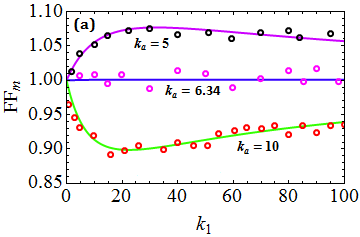} \includegraphics[width=5.5cm,height=3.5cm]{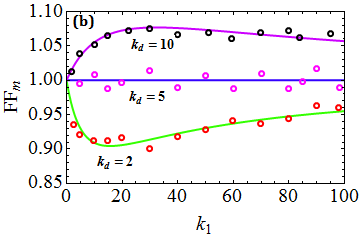}
\par\end{centering}
\caption{\label{fig:TS-Fano-K1}Plot of Fano factor with the variation of $k_{1}$with
parameter (a) $k_{a}$ and (b) $k_{d}$. The other rate constants
are $J_{m}=10,\,k_{m}=1$, $k_{2}=1,$ $k_{d}=10$ in figure (a) and
$k_{a}=5$ in figure (b). The solid lines are drawn from analytical
expression (Eq. \ref{eq:TS-8}) and hollow circles are obtained from
the stochastic simulation based on the Gillespie algorithm. }
\end{figure}

The variation of mean mRNA levels for both the scenarios, with and
without reinitiation, are plotted with the rate constants $k_{a}$,
$k_{d}$ and $J_{m}$ in figure \ref{fig:Variation-of-mean Two-state}.
We see from the figures that the reinitiation of transcription helps
to keep the mean mRNA levels at higher values for lower values of
$k_{a}$ and $J_{m}$ and for almost all values of $k_{d}$. Fano
factor remains lower for all values of $k_{a}$, $k_{d}$ and $J_{m}$
(Figs. \ref{fig:Plot-of-Fano-Two-state}(a), \ref{fig:Plot-of-Fano-Two-state}(b)
and \ref{fig:Plot-of-Fano-Two-state}(c)). If one considers pre-initiation
and initiation complexes in the transcriptional reinitiation process
\cite{key-35} rather than only the initiation complex then the Fano
factor further reduces below unity (Appendix B). Fano factor can also
be higher due to the transcriptional reinitiation process with other
sets of rate constants \cite{key-17,key-39}. The Fano factor can
have three different phases, Poissonian ($FF=1)$, super-Poissonian
($FF>1$) and sub-Poissonian ($FF<1$), when plotted against $k_{1}$
with $k_{a}$, $k_{d}$ and $J_{m}$ as parameters as shown in figure
\ref{fig:TS-Fano-K1}(a), figure \ref{fig:TS-Fano-K1}(b) and in ref.
\cite{key-39} respectively. The rate constants $k_{a}=6.34$ and
$k_{d}=5$ in figure \ref{fig:TS-Fano-K1}(a) and figure \ref{fig:TS-Fano-K1}(b)
respectively can be considered as the critical value of that rate
constants as that values sharply divide the super-Poissonian and sub-Poissonian
Fano factor regimes \cite{key-39}. 

The expression of mean mRNA level (Eq.(\ref{eq:TS-7})) can be written
as

\begin{equation}
<m^{tswr}>=\frac{<m^{tswtr}>}{1-\beta_{1}<m^{tswtr}>+\beta_{2}<m^{tswtr}>}\label{eq:TS-10}
\end{equation}

where $\beta_{1}=\frac{k_{d}\,k_{m}}{k_{a\,}J_{m}}$ and $\beta_{2}=\frac{(k_{a}+\,k_{d})\,(J_{m}+k_{2})\,k_{m}}{k_{a\,}k_{1}J_{m}}$. 

The expression of Fano factor (Eq. (\ref{eq:TS-8})) can also be expressed
as

\begin{equation}
FF_{m}^{tswr}=1-\gamma_{1}<m^{tswtr}>+\gamma_{2}<m^{tswtr}>\label{eq:TS-11}
\end{equation}

where $\gamma_{1}=\frac{k_{m}k_{1}(k_{a}+k_{d})\,a_{1}}{(a_{1}k_{m}+a_{2})\,a_{2}}$and
$\gamma_{2}=\frac{k_{m}k_{1}(k_{a}+k_{d})}{(a_{1}k_{m}+a_{2})\,k_{a\,}}$.

In general, for $\beta_{1}=0$ and $\beta_{2}\neq0$ ($\beta_{2}=0$
and $\beta_{1}\neq0$), the expression for mean in equation (\ref{eq:TS-10})
looks like the expression of the gain with linear negative (positive)
feedback network in electronic circuit with $\beta_{2}$ ($\beta_{1}$)
as the feedback factor. For the non-zero value of $\beta_{1}$ and
$\beta_{2}$, the expression for mean in equation (\ref{eq:TS-10})
can be considered as the mean mRNA from a network with mixed feedback
i.e., both positive and negative feedback. Therefore, $\beta_{1}$($\beta_{2}$)
is working here as the positive (negative) feedback factor. For $\beta_{1}>\beta_{2}$
($\beta_{1}<\beta_{2}$) the positive (negative) feedback nature dominates
and the mean mRNA level with reinitiation ($<m^{tswr}>$) becomes
higher (lower) than that without reinitiation process ($<m^{tswtr}>$).
Again, as far as the Fano factor is concerned, the positive (negative)
feedback nature dominates for $\gamma_{2}>\gamma_{1}$ ($\gamma_{2}<\gamma_{1}$).
The expression of mean mRNA (Eq. (\ref{eq:TS-10})) and Fano factor
(Eq. (\ref{eq:TS-11})) shows that the two-state gene regulatory network
with reinitiation of transcription (Fig. \ref{fig:Two-state}(b))
can behave as mixed feedback network.

The mean mRNA level and Fano factor can be higher or lower due to
the reinitiation of transcription compared to the two-state gene expression
process without reinitiation. From equation (\ref{eq:TS-2}) and (\ref{eq:TS-7})
or from equation (\ref{eq:TS-10}), we have the condition of higher
average mRNA level in presence of reinitiation of transcription as

\begin{equation}
\beta_{1}>\beta_{2}\;\quad or\;(J_{m}+k_{2})<\frac{k_{d}\,k_{1}}{k_{a}+k_{d}}\label{eq:TS-12}
\end{equation}

From equation (\ref{eq:TS-8}) or (\ref{eq:TS-11}), we have the condition
of sub-Poissonian Fano factor as \cite{key-39}

\begin{equation}
\gamma_{1}>\gamma_{2}\;\quad or\;(J_{m}+k_{2})<\frac{k_{a}}{k_{d}}(k_{a}+k_{d}+k_{m})\label{eq:TS-13}
\end{equation}

\begin{figure}[H]
\begin{centering}
\includegraphics[width=5cm,height=3cm]{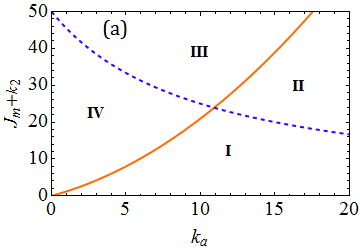}$ $
\includegraphics[width=5cm,height=3cm]{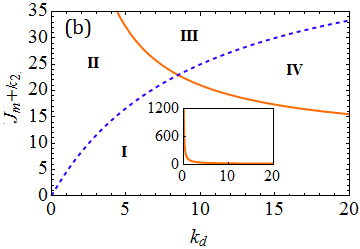} 
\par\end{centering}
\caption{\label{fig:four regions in two-state}The plot of ($J_{m}+k_{2})$
versus (a) $k_{a}$ and (b) $k_{d}$ shows the four different regions.
The rate constants are taken as $k_{1}=50$ and $k_{m}=1$and $k_{d}=10$
(in (a)), and $k_{a}=10$ (in (b)). The condition given in equation
(\ref{eq:TS-12}) is satisfied in Regions I and IV whereas the condition
given in equation (\ref{eq:TS-13}) is satisfied in Regions I and
II.}
\end{figure}

Two conditions in equations (\ref{eq:TS-12}) and (\ref{eq:TS-13})
divide the whole permissible space in ($k_{a},\:J_{m}+k_{2}$) and
($k_{d},\:J_{m}+k_{2}$) space into four different regions with different
conditions of Fano factor and mean mRNA level. The four regions are
identified as: Region \textbf{I}: $FF_{m}^{tswr}<1$ and $r>1$; Region
\textbf{II}: $FF_{m}^{tswr}<1$ and $r<1$; Region \textbf{III}: $FF_{m}^{tswr}>1$
and $r<1$; Region \textbf{IV}: $FF_{m}^{tswr}>1$ and $r>1$; where
$r=<m^{tswr}>/<m^{tswtr}>$ and shown in figure (\ref{fig:four regions in two-state}).
The mean and Fano factor of mRNA can be in any one of the four regions
depending on the rate constants $k_{a},\:\:k_{d},\:k_{1},$ $J_{m}+k_{2}$
and $k_{m}$. The rate constants $k_{a}$ and $k_{d}$ are generally
function of transcription factors and therefore, can be modulated
\cite{key-17,key-18,key-39D}. Thus the mean and Fano factor at the
mRNA level can be changed according to the cellular requirement by
changing the number of transcription factors in the cell.

In the two-state process (Fig. \ref{fig:Two-state}(a)) Fano factor
is always greater than unity and there is a specific mean mRNA level
depending on the rate constants $k_{a}$, $k_{d}$, $J_{m}$ and $k_{m}$.
But, as reinitiation of transcription is added in the two-state gene
expression process, we get four different options on Fano factor and
mean mRNA in the $(k_{a},J_{m}+k_{2})$ (Fig. \ref{fig:four regions in two-state}(a))
or $(k_{d},J_{m}+k_{2})$ (Fig. \ref{fig:four regions in two-state}(b))
space. Four distinct regions in the parameter space are obtained by
two intersecting curves corresponding to the two conditions given
in equations (\ref{eq:TS-12}) and (\ref{eq:TS-13}). The intersecting
point of the two curves gives the condition $r=1$ and $FF_{m}^{tswr}=1$
simultaneously. With respect to the feedback features of a network,
we see that the gene regulatory network ((Fig. \ref{fig:Two-state}(b)))
behaves like a network with inherent negative (positive) feedback
in Region II (Region IV). In Regions I and III, the two-state network
with reinitiation of transcription behaves like a gene regulation
with inherent mixed feedback. The mixed feedback nature is of two
types: either it behaves as negative feedback for the mean ($r<1$)
and positive feedback for Fano factor ($FF_{m}^{tswr}>1$) or it behaves
as positive feedback for mean ($r>1$) and negative feedback for Fano
factor ($FF_{m}^{tswr}<1$).

\begin{figure}[H]
\begin{centering}
\includegraphics[width=5.5cm,height=3.5cm]{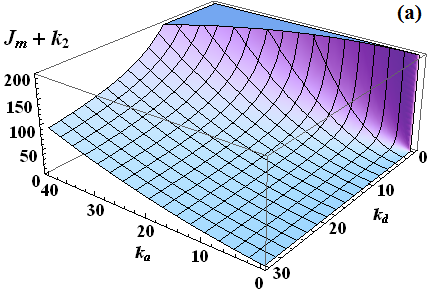} \includegraphics[width=5.5cm,height=3.5cm]{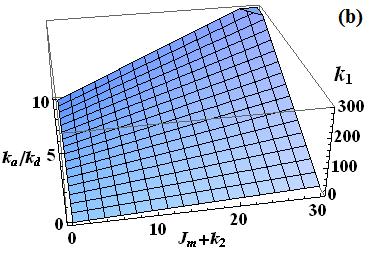}
\par\end{centering}
\begin{centering}
\includegraphics[width=5.5cm,height=3.5cm]{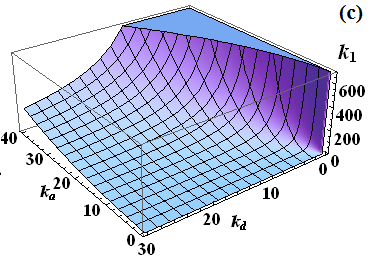} \includegraphics[width=5.5cm,height=3.5cm]{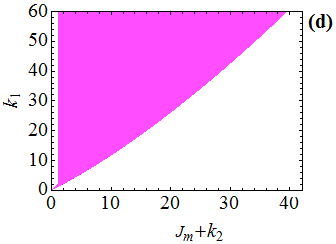}
\par\end{centering}
\caption{\label{fig:TS-Surface plot}(a) The surface plot for the ($J_{m}+k_{2})$
with independent variables $k_{a}$ and $k_{d}$. The critical condition
$FF_{m}=1$ is satisfied on that surface. (b) The surface plot for
the parameter $k_{1}$ with independent variables ($J_{m}+k_{2})$
and $k_{a}/k_{d}$. The condition $r=1$ is satisfied on that surface.
(c) The surface plot for the rate constant $k_{1}$ with independent
variables $k_{a}$ and $k_{d}$. On that surface, both the conditions
$r=1$ and $FF_{m}^{tswr}=1$ are satisfied provided the ($J_{m}+k_{2})$
is chosen from the surface in (a) with same $k_{a}$ and $k_{d}$.
(d) Parametric plot for $k_{1}$ versus ($J_{m}+k_{2})$ with $k_{a}$
and $k_{d}$ as parameters. We use $k_{m}=1$ for all plots.}
\end{figure}

Now, we can express $k_{1}$ in terms of $k_{a}$, $k_{d}$ and $k_{m}$
with the equality conditions in equations (\ref{eq:TS-12}) and (\ref{eq:TS-13}),
as

\begin{equation}
k_{1}=\frac{k_{a}}{k_{d}^{2}}(k_{a}+k_{d})(k_{a}+k_{d}+k_{m})\label{eq:TS-14}
\end{equation}

The equation (\ref{eq:TS-13}) with the equality sign gives the critical
condition ($FF_{m}^{tswr}=1$) and that condition is satisfied at
all the points on the surface in figure \ref{fig:TS-Surface plot}(a).
The equation (\ref{eq:TS-12}) with the equality sign gives the condition
$r=1$ and all the points on the surface in figure \ref{fig:TS-Surface plot}(b)
satisfy that condition (The ratio $k_{a}/k_{d}$ is chosen as independent
variable to avoid the assignment of constant value to any one of them).
Both the conditions, $r=1$ and $FF_{m}^{tswr}=1$, are satisfied
at all the points on the surface in figure \ref{fig:TS-Surface plot}(c)
provided the value of  ($J_{m}+k_{2})$ is chosen from the surface
of figure \ref{fig:TS-Surface plot}(a) for the same $k_{a}$ and
$k_{d}$ as in \ref{fig:TS-Surface plot}(c). For a given $k_{a}$,
$k_{d}$ and $k_{m}$, the value of the rate constants $k_{1}$ from
equation (\ref{eq:TS-14}) and ($J_{m}+k_{2})$ from equation (\ref{eq:TS-13})
(with equality) can be obtained to have both the conditions $r=1$
and $FF_{m}^{tswr}=1$. The value of rate constants $k_{1}$ and ($J_{m}+k_{2})$
in the pink shaded region of the plot in figure \ref{fig:TS-Surface plot}(d)
satisfy the conditions $r=1$ and $FF_{m}^{tswr}=1$ simultaneously.

\subsection{Two-state gene expression (Suter model) without and with reinitiation }

\begin{figure}[H]
\begin{centering}
\includegraphics[width=5cm,height=2cm]{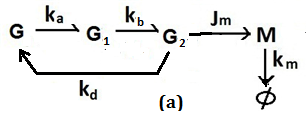}
\includegraphics[width=5cm,height=2cm]{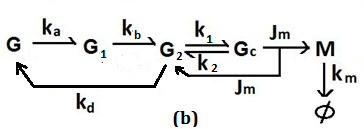}
\par\end{centering}
\caption{\label{fig:Suter Model}Reaction scheme with rate constants for two-state
gene expression (Suter model) (a) without and (b) with reinitiation.
$k_{b}$ ($k_{d}$) is the activation (deactivation) rate constant
and $k_{a}$ is the rate constant for transition from $G$ to $G_{1}$.
$k_{1}$ is the rate constant of initiation complex formation and
$k_{2}$ is the rate constant of dissociation of RNAP-II from initiation
complex. $J_{m}$ is the transcription rate constant and $k_{m}$
is the mRNA degradation rate constant. }
\end{figure}

The regulated gene expression is an important and essential property
of a complex cellular system. Though many experimental results are
modeled with the two-state process, Suter et al \cite{key-27} observe
something different in the mammalian system. They observe gamma-distributed
off time in gene regulation rather than the exponentially distributed
off time in the two-state process. Suter et al model their experimental
observation by a gene regulatory network shown in figure \ref{fig:Suter Model}(a).
In their model network, the gene can be in three possible states,
one active ($G_{2}$) and two inactive states ($G$ and $G_{1}$)
and random transitions take place between the states according to
the reaction scheme in figure \ref{fig:Suter Model}(a). The mRNA
synthesis takes place only from the active state of the gene with
rate constant $J_{m}$. 

The Master equation corresponding to the figure \ref{fig:Suter Model}(a)
is given by

\begin{equation}
\begin{array}{ccc}
\frac{\partial p(n_{1},n_{2},n_{3},t)}{\partial t} & = & k_{a}[(l-n_{1}-n_{2}+1)p(n_{1}-1,n_{2},n_{3},t)-(l-n_{1}-n_{2})p(n_{1},n_{2},n_{3,}t)]\\
 &  & +k_{b}[(n_{1}+1)p(n_{1}+1,n_{2}-1,n_{3},t)-n_{1}p(n_{1},n_{2},n_{3,}t)]\\
 &  & +k_{d}[(n_{2}+1)p(n_{1},n_{2}+1,n_{3},t)-n_{2}p(n_{1},n_{2},n_{3,}t)]\\
 &  & +J_{m}[n_{2}p(n_{1},n_{2},n_{3}-1,t)-n_{2}p(n_{1},n_{2},n_{3,}t)]\\
 &  & +k_{m}[(n_{3}+1)p(n_{1},n_{2},n_{3}+1,t)-n_{3}p(n_{1},n_{2},n_{3,}t)]
\end{array}\label{eq:Su-1}
\end{equation}

The expression for mean and Fano factor at mRNA level corresponding
to figure \ref{fig:Suter Model}(a) are given by (for $l=1$)

\begin{equation}
<m^{suwtr}>=\frac{J_{m}k_{a}k_{b}}{(k_{a}k_{b}+k_{a}k_{d}+k_{b}k_{d})k_{m}}=\frac{J_{m}k_{a}k_{b}}{C_{1}k_{m}}\label{eq:Su-2}
\end{equation}

\begin{equation}
FF_{m}^{suwtr}=1+\frac{J_{m}k_{d}[(k_{b}+k_{m})(k_{a}+k_{b})+k_{a}^{2}]}{\begin{array}{c}
\{k_{a}(k_{b}+k_{d})+k_{b}k_{d}\}\{(k_{b}+k_{m})(k_{d}+k_{m})+k_{a}(k_{b}+k_{d}+k_{m})\}\end{array}}=1+\frac{J_{m}k_{d}[(k_{b}+k_{m})(k_{a}+k_{b})+k_{a}^{2}]}{C_{1}\,C_{2}}\label{eq:Su-3}
\end{equation}

where $C_{1}=k_{a}(k_{b}+k_{d})+k_{b}k_{d}$ and $C_{2}=(k_{b}+k_{m})(k_{d}+k_{m})+k_{a}(k_{b}+k_{d}+k_{m})$. 

Now let us consider the gene transcriptional regulatory network with
the reinitiation of transcription by RNAP-II (Fig. \ref{fig:Suter Model}(b)).
We have the Master equation corresponding to the figure \ref{fig:Suter Model}(b)
as

\begin{equation}
\begin{array}{ccc}
\frac{\partial p(n_{1},n_{2},n_{3},n_{4},,t)}{\partial t} & = & k_{a}[\{l-(n_{1}-1+n_{2}+n_{3}\}p(n_{1}-1,n_{2},n_{3},n_{4},t)\\
 &  & -\{l-(n_{1}+n_{2}+n_{3})\}p(n_{1},n_{2},n_{3,}n_{4},t)]\\
 &  & +k_{b}[(n_{1}+1)p(n_{1}+1,n_{2}-1,n_{3},n_{4},t)-n_{1}p(n_{1},n_{2},n_{3,}n_{4},t)]\\
 &  & +k_{d}[(n_{2}+1)p(n_{1},n_{2}+1,n_{3},n_{4},t)-n_{2}p(n_{1},n_{2},n_{3,}n_{4},t)]\\
 &  & +k_{1}[(n_{2}+1)p(n_{1},n_{2}+1,n_{3}-1,n_{4},t)-n_{2}p(n_{1},n_{2},n_{3,}n_{4},t)]\\
 &  & +k_{2}[(n_{3}+1)p(n_{1},n_{2}-1,n_{3}+1,n_{4}-1,t)-n_{3}p(n_{1},n_{2},n_{3,}n_{4},t)]\\
 &  & +J_{m}[(n_{3}+1)p(n_{1},n_{2}-1,n_{3}+1,n_{4}-1,t)-n_{3}p(n_{1},n_{2},n_{3,}n_{4},t)]\\
 &  & +k_{m}[(n_{4}+1)p(n_{1},n_{2},n_{3},n_{4}+1,t)-n_{4}p(n_{1},n_{2},n_{3,}n_{4},t)]
\end{array}\label{eq:Su-4}
\end{equation}

The mean mRNA and Fano factor corresponding to figure \ref{fig:Suter Model}(b)
are given by (for $l=1$)

\begin{equation}
<m^{suwr}>=\frac{J_{m}k_{1}k_{a}k_{b}}{[k_{a}k_{b}(J_{m}+k_{1}+k_{2})+(J_{m}+k_{2})(k_{a}k_{d}+k_{b}k_{d})]k_{m}}\label{eq:Su-5}
\end{equation}

\begin{equation}
FF_{m}^{suwr}=1-\frac{J_{m}k_{1}k_{a}k_{b}}{[k_{1}k_{a}k_{b}+(J_{m}+k_{2})C_{1}]k_{m}}+\frac{J_{m}C_{3}}{[C_{3}+(J_{m}+k_{2}+k_{m})C_{2}]k_{m}}\label{eq:Su-6}
\end{equation}

where $C_{3}=k_{1}(k_{a}+k_{m})(k_{b}+k_{m})$,

\begin{figure}[H]
\begin{centering}
\includegraphics[width=5.5cm,height=3.5cm]{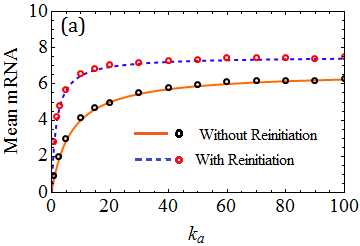} \includegraphics[width=5.5cm,height=3.5cm]{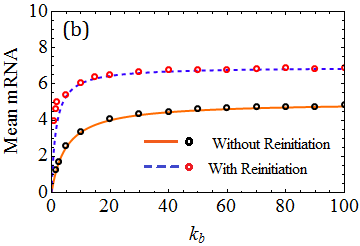}
\par\end{centering}
\begin{centering}
\includegraphics[width=5.5cm,height=3.5cm]{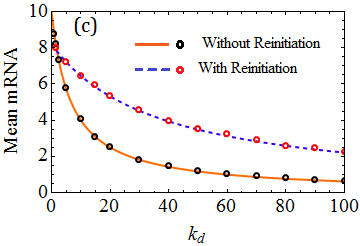} \includegraphics[width=5.5cm,height=3.5cm]{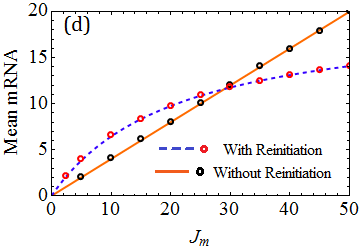}
\par\end{centering}
\caption{\label{fig:Su-Variation-of-mean}Variation of mean mRNA with (a) $k_{a}$,
(b) $k_{b}$, (c) $k_{d}$, and (d) $J_{m}$. The dashed (solid) lines
are from exact analytical calculations corresponding to equation (\ref{eq:Su-5})
(Eq. (\ref{eq:Su-2})). The hollow circles are generated from the
stochastic simulation based on the Gillespie algorithm. The rate constants
are $k_{1}=50,$ $k_{2}=1,$ $k_{m}=1$. In figure (a) $k_{d}=10$
$k_{b}=20$ and $J_{m}=10$. In figure (b) $k_{a}=10,$ $k_{d}=10$
and $J_{m}=10$ . In figure (c) $k_{a}=10,$ $k_{b}=20$ and $J_{m}=10$.
In figure (d) $k_{a}=10,$ $k_{d}=10$ and $k_{b}=20$. }
\end{figure}

Figures \ref{fig:Su-Variation-of-mean}(a), \ref{fig:Su-Variation-of-mean}(b),
\ref{fig:Su-Variation-of-mean}(c) and \ref{fig:Su-Variation-of-mean}(d)
show the variation of mean mRNA number with the rate constants $k_{a}$,
$k_{b}$, $k_{d}$ and $J_{m}$ respectively. All figures show that
the reinitiation of transcription favours the higher mean mRNA levels
with reinitiation of transcription. 

\begin{figure}[H]
\begin{centering}
\includegraphics[width=5.5cm,height=3.5cm]{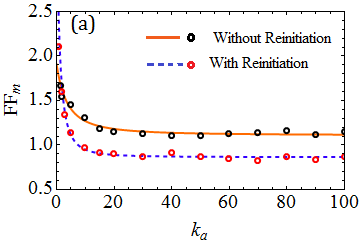}
\includegraphics[width=5.5cm,height=3.5cm]{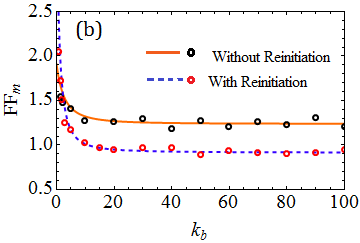}
\par\end{centering}
\begin{centering}
\includegraphics[width=5.5cm,height=3.5cm]{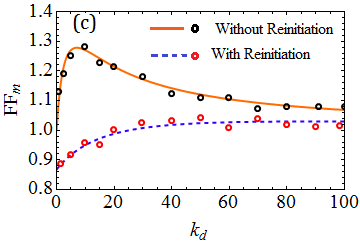}\includegraphics[width=5.5cm,height=3.5cm]{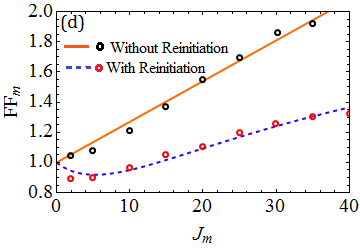} 
\par\end{centering}
\caption{\label{fig:Su-Variation-of-Fano}Variation of Fano factor at mRNA
level ($FF_{m})$ with different conditions. Dashed (solid) lines
are drawn from analytical calculation for figure \ref{fig:Suter Model}(b)
(Fig. \ref{fig:Suter Model}(a)). The hollow circles are generated
from the stochastic simulation based on the Gillespie algorithm. The
rate constants are $k_{1}=50,$ $k_{2}=1,$ $k_{m}=1$. In figure
(a) $k_{d}=10,$ $k_{b}=20$ and $J_{m}=10$. In figure (b) $k_{a}=10,$
$k_{d}=10$ and $J_{m}=10$ . In figure (c) $k_{a}=10,$ $k_{b}=20$
and $J_{m}=10$. In figure (d) $k_{a}=10,$ $k_{d}=10$ and $k_{b}=20$.}
\end{figure}

\begin{figure}[H]
\begin{centering}
\includegraphics[width=5.5cm,height=3.5cm]{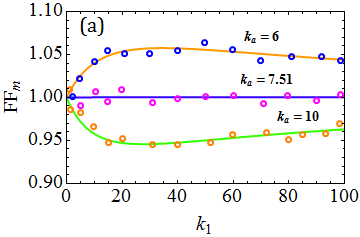} \includegraphics[width=5.5cm,height=3.5cm]{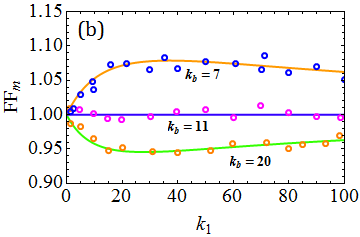}
\par\end{centering}
\begin{centering}
\includegraphics[width=5.5cm,height=3.5cm]{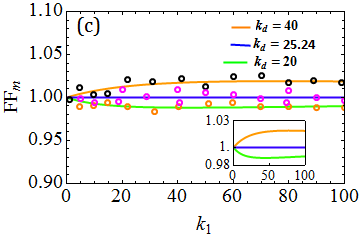} \includegraphics[width=5.5cm,height=3.5cm]{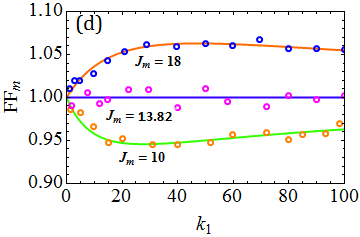}
\par\end{centering}
\caption{\label{fig:Su-FFm-K1}Variation of Fano factor with the rate constant
$k_{1}$ with (a) $k_{a}$ (b) $k_{b}$ (c) $k_{d}$ and (d) $J_{m}$
as parameter. The solid lines are drawn from analytical expression
(Eq. \ref{eq:Su-6}) and hollow circles are obtained from the stochastic
simulation based on the Gillespie algorithm. The rate constants are
taken as $k_{d}=10,$ $k_{b}=20$, $k_{1}=50,$ $k_{2}=1,$ $J_{m}=10$
and $k_{m}=1$ in (a). The rate constants are $k_{a}=10,$ $k_{d}=10$,
$k_{1}=50,$ $k_{2}=1,$ $J_{m}=10$ and $k_{m}=1$. in (b). In (c),
the rate constants are $k_{a}=10,$ $k_{b}=20$, $k_{1}=50,$ $k_{2}=1,$
$J_{m}=10$ and $k_{m}=1$. In (d), the rate constants are $k_{a}=10,$
$k_{d}=10,$ $k_{b}=20$, $k_{1}=50,$ $k_{2}=1,$ and $k_{m}=1$. }
\end{figure}

The variation of Fano factors are plotted with the rate constants
$k_{a}$, $k_{b}$, $k_{d}$ and $J_{m}$ respectively in figures
\ref{fig:Su-Variation-of-Fano}(a) - \ref{fig:Su-Variation-of-Fano}(d).
The figures show that the Fano factor is always lower with the reinitiation
of gene transcription. The variation of the Fano factor with the rate
constant $k_{1}$ is shown in figure \ref{fig:Su-FFm-K1}. The figures
also show that the three different regimes of Fano factor viz., Poissonian,
sub-Poissonian and super-Poissonian, are likely to occur with the
reinitiation of transcription. That is a unique feature of the Fano
factor for a gene regulatory network with reinitiation of transcription.
In the Suter model, $k_{b}$ is an extra parameter by which the mean
and Fano factor can be controlled. It can be shown that for higher
$k_{b}$ ($k_{b}>30$) the Suter model merges to the two-state model. 

The expression of mean mRNA (Eq. (\ref{eq:Su-5})) can also be expressed
as

\begin{equation}
<m^{suwr}>=\frac{<m^{suwtr}>}{1-\delta_{1}<m^{suwtr}>+\delta_{2}<m^{suwtr}>}\label{eq:Su-7}
\end{equation}

where $\delta_{1}=\frac{k_{d}\,k_{m}(k_{a}+\,k_{b})}{k_{a\,}k_{b}J_{m}}$
and $\delta_{2}=\frac{C_{1}(J_{m}+k_{2})\,k_{m}}{k_{a\,}k_{1}J_{m}k_{b}}$.
Here $\delta_{1}$($\delta_{2}$) is working as the positive (negative)
feedback factor. For $\delta_{1}>\delta_{2}$ ($\delta_{1}<\delta_{2}$)
the positive (negative) feedback nature dominates and the mean mRNA
level with reinitiation ($<m^{suwr}>$) becomes higher (lower) than
that without reinitiation process ($<m^{suwtr}>$).

The expression of Fano factor (Eq. (\ref{eq:TS-8})) can also be expressed
as

\begin{equation}
FF_{m}^{suwr}=1-B_{1}<m^{suwtr}>+B_{2}<m^{suwtr}>\label{eq:Su-8}
\end{equation}

where $B_{1}=\frac{(J_{m}+k_{2})}{k_{1}}+\frac{k_{a}\,k_{b}}{C_{1}}$
and $B_{2}=\frac{C_{3}\,C_{1}}{\{C_{3}+(J_{m}+k_{2}+k_{m})C_{2}\}\,k_{a\,}k_{b}}$.

Again, as far as the Fano factor is concerned, the positive (negative)
feedback nature dominates for $B_{2}>B_{1}$ ($B_{2}<B_{1}$). The
expressions of mean mRNA (Eq. (\ref{eq:Su-7})) and Fano factor (Eq.
(\ref{eq:Su-8})) show that the gene regulatory network following
Suter model with reinitiation of transcription (Fig. \ref{fig:Suter Model}(b))
can also behave as mixed feedback network.

We see that the Fano factor is reduced by the transcriptional reinitiation
process as observed in two-state also. From the equations (\ref{eq:Su-7})
and (\ref{eq:Su-8}) we find that the average mRNA level can be greater
with the reinitiation in gene transcription provided

\begin{equation}
(J_{m}+k_{2})<\frac{k_{1}k_{d}\,(k_{a}+k_{b})}{(k_{a}k_{b}+k_{a}k_{d}+k_{b}k_{d})}\label{eq:Su-9}
\end{equation}

From equation (\ref{eq:TS-8}), we have the condition of sub-Poissonian
Fano factor as \cite{key-39}

\begin{equation}
(J_{m}+k_{2})<\frac{k_{a}k_{b}}{k_{d}}\frac{\{(k_{b}+k_{m})(k_{d}+k_{m})+k_{a}(k_{b}+k_{d}+k_{m})\}}{(k_{a}^{2}+k_{b}^{2}+k_{a}k_{b}+k_{a}k_{m}+k_{b}k_{m})}\label{eq:Su-10}
\end{equation}

\begin{figure}[H]
\begin{centering}
\includegraphics[width=5.5cm,height=3.5cm]{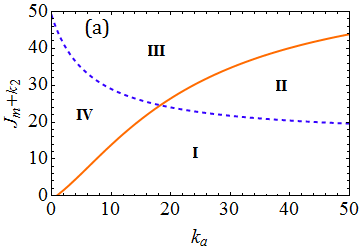}$\:\:$\includegraphics[width=5.5cm,height=3.5cm]{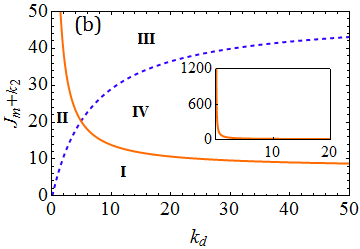}
\par\end{centering}
\caption{\label{fig:Su-Phase diagram}The plot of ($J_{m}+k_{2})$ versus (a)
$k_{a}$ and (b) $k_{d}$ shows the four different regions. The rate
constants are taken as $k_{b}=20,$ $k_{1}=50$, $k_{m}=1,$ $k_{d}=10$
(in (a)), and $k_{a}=10$ (in (b)). The condition given in equation
(\ref{eq:Su-9}) is satisfied in regions I and IV whereas the condition
given in equation (\ref{eq:Su-10}) is satisfied in regions I and
II. }
\end{figure}

Here also, two conditions in equations (\ref{eq:Su-9}) and (\ref{eq:Su-10})
divide the whole permissible space in ($k_{a},\:J_{m}+k_{2}$) and
($k_{d},\:J_{m}+k_{2}$) space into four different regions with different
conditions of Fano factor and mean mRNA levels. The four regions are
identified as: Region \textbf{I}: $FF_{m}^{suwr}<1$ and $s>1$; Region
\textbf{II}: $FF_{m}^{suwr}<1$ and $s<1$; Region \textbf{III}: $FF_{m}^{suwr}>1$
and $s<1$; Region \textbf{IV}: $FF_{m}^{suwr}>1$ and $s>1$; where
$s=<m^{suwr}>/<m^{suwtr}>$ and shown in figure (\ref{fig:Su-Phase diagram}).
The mean and Fano factor of mRNA can be in any one of the four regions
depending on the rate constants $k_{a},\:k_{b},\:k_{d},\:k_{1},$
$J_{m}+k_{2}$ and $k_{m}$.

Figure \ref{fig:Su-Phase diagram} shows the four different regions
in the $(k_{a},\,J_{m}+k_{2})$ (Fig. \ref{fig:Su-Phase diagram}(a))
or $(k_{d},\,J_{m}+k_{2})$ (Fig. \ref{fig:Su-Phase diagram}(b))
space in Suter model. With the gradual decrease in the rate constant
$k_{b}$, the area of the Region II decreases gradually and the two
curves intersect at higher (lower) value of $k_{a}$ ($k_{d}$) in
figure \ref{fig:Su-Phase diagram}(a) (Fig. \ref{fig:Su-Phase diagram}(b)).
A cell can choose any one of the four regions according to its functional
requirement by modulating the rate constants $k_{a}$, $k_{d}$ or
$k_{b}$.

\begin{figure}[H]
\begin{centering}
\includegraphics[width=5.5cm,height=3.5cm]{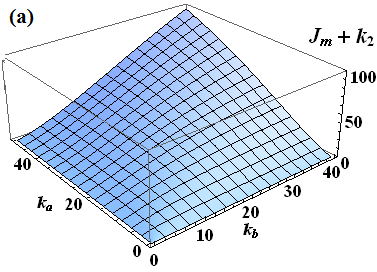} \includegraphics[width=5.5cm,height=3.5cm]{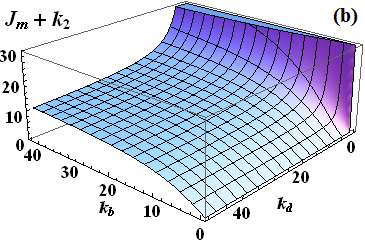}\includegraphics[width=5.5cm,height=3.3cm]{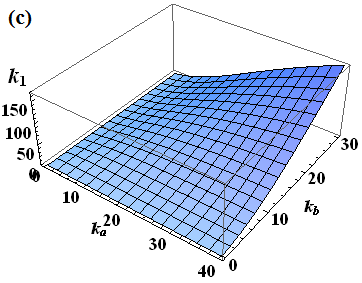}
\par\end{centering}
\begin{centering}
\includegraphics[width=5.5cm,height=3.5cm]{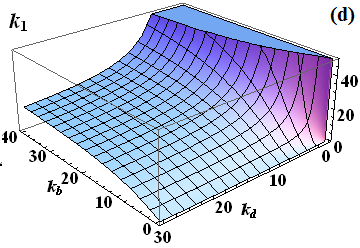} \includegraphics[width=5.5cm,height=3.5cm]{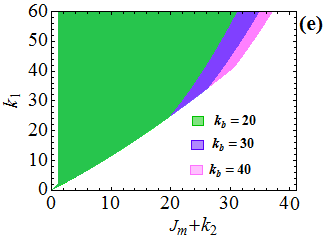}
\par\end{centering}
\caption{\label{fig:SU-Surface plot}The surface plots for the ($J_{m}+k_{2})$
with variables (a) $k_{a}$ and $k_{b}$ (b) $k_{b}$ and $k_{d}$
with the equality condition in equation (\ref{eq:Su-10}). For the
surface in (a) $k_{a}=10$ and in (b) $k_{d}=10$. The surface plots
for the rate constant $k_{1}$ with variables (c) $k_{a}$, $k_{b}$
and (d) $k_{b}$, $k_{d}$ for the simultaneous condtions for mean
and Fano factor ($r=1$ and $FF_{m}^{tswr}=1$) provided the ($J_{m}+k_{2})$
are chosen from (a) and (b) respectively for the same independent
variables. (e) Parametric plot for $k_{1}$ versus ($J_{m}+k_{2})$
with $k_{a}$ and $k_{d}$ as parameters for different $k_{b}$. We
use $k_{m}=1$ for all plots.}
\end{figure}

From equations (\ref{eq:Su-9}) and (\ref{eq:Su-10}) (with equality
sign in both the equations), we can have the rate constant $k_{1}$
in terms of $k_{a}$, $k_{b}$, $k_{d}$ and $k_{m}$ as

\begin{equation}
k_{1}=\frac{k_{a}k_{b}(k_{a}k_{b}+k_{a}k_{d}+k_{b}k_{d})}{k_{d}^{2}\,(k_{a}+k_{b})}\frac{\{(k_{b}+k_{m})(k_{d}+k_{m})+k_{a}(k_{b}+k_{d}+k_{m})\}}{(k_{a}^{2}+k_{b}^{2}+k_{a}k_{b}+k_{a}k_{m}+k_{b}k_{m})}\label{eq:SU-11}
\end{equation}

The equation (\ref{eq:Su-10}) with the equality sign gives the critical
condition ($FF_{m}^{suwr}=1$) and that condition is satisfied at
all the points on the surface in figure \ref{fig:SU-Surface plot}(a)
and (b). The equation (\ref{eq:Su-9}) with the equality sign gives
the condition $r=1$. Both the conditions, $r=1$ and $FF_{m}^{tswr}=1$,
are satisfied at all the points on the surfaces in figure \ref{fig:SU-Surface plot}(c)
and (d) provided the value of ($J_{m}+k_{2})$ is chosen from the
surface of figure \ref{fig:SU-Surface plot}(a) and (b) respectively
for the independent variables $k_{a}$, $k_{b}$ and $k_{d}$. For
a given $k_{a}$, $k_{b}$ $k_{d}$ and $k_{m}$, the rate constants
$k_{1}$ from equation (\ref{eq:SU-11}) and ($J_{m}+k_{2})$ from
equation (\ref{eq:Su-10}) (with equality) can be obtained to have
both the conditions $r=1$ and $FF_{m}^{tswr}=1$. Figure \ref{fig:SU-Surface plot}(e)
show the parametric plot for $k_{1}$ versus ($J_{m}+k_{2})$ with
$k_{a}$ and $k_{d}$ as parameters for different $k_{b}$. For low
$k_{b}$, the shaded area is lower than that compared to the figure
\ref{fig:TS-Surface plot}(d).

\section{Conclusion}

In this article, we study the effect of transcriptional reinitiation
by RNAP-II in gene expression. Transcriptional reinitiation is an
important step in gene expression though it is ignored in most of
the model networks assuming it has an insignificant role in mRNA and
protein levels. But, Blake et. el. identify that reinitiation of transcription
can be crucial in eukaryotic systems \cite{key-17,key-18}. To find
out the effect of transcriptional reinitiation on phenotypic variability,
we consider different gene regulatory networks, with and without the
reinitiation step. We find the analytical expression of mean and Fano
factor at mRNA level at steady state for constitutive, two-state,
and Suter model. We compare our analytical results with the results
obtained from stochastic simulation using the Gillespie algorithm.
In this work, the rate constants are chosen from different works \cite{key-21,key-39}.
When the constitutive gene network is analyzed in presence of reinitiation,
we observe that the mean mRNA level and Fano factor both are reduced.
That is happened due to the reduction of the effective transcription
rate of mRNA synthesis. The behaviour is similar to a negative feedback
amplifier which reduces the gain and noise. Though, there is a fundamental
difference between a negative feedback amplifier in electronic circuits
and the observed negative feedback like behavior in the constitutive
gene network with transcriptional reinitiation. In the electronic
negative feedback circuit, a fraction of the output voltage is fed
back to the input but in the present reinitiation-based gene regulatory
circuit the gene product (mRNAs or proteins) is not involved at all
in the synthesis process. Thus the reinitiation based negative feedback
in constitutive gene transcription is entirely inherent in nature.
Again, with the non-zero rate constant of reverse reaction i.e., with
the unsuccessful transcription from the initiation complex, the average
mRNA level further decreases but the Fano factor increases than before
but still remains in the sub-Poissonian region. Thus, even with the
reverse reaction, the reinitiation of transcription in the constitutive
model has the capability to decrease the Fano factor below unity. 

Then we study the two-state gene expression model with transcriptional
reinitiation. In two-state model without the reinitiation the Fano
factor at mRNA level is higher than unity due to random transitions
between the active and inactive states of the gene. Now, with the
reinitiation of transcription in the two-state model, we observe four
different phenotypic outcomes ( $r<1$ and $FF_{m}^{tswr}<1$; $r>1$
and $FF_{m}^{tswr}>1$; $r<1$ and $FF_{m}^{tswr}>1$; $r>1$ and
$FF_{m}^{tswr}<1$; where $r=<m^{tswr}>/<m^{tswtr}>$) depending on
the rate constants of the biochemical reactions. Similar behaviour
is observed for the Suter model also. Though the mean mRNA level is
higher and the Fano factor is lower over a wide region of parameter
variation in the Suter model. We find that the gene regulatory network
like the two-state and Suter model with RNAP-II based transcriptional
reinitiation process can behave as either the inherent negative ($r<1$
and $FF_{m}^{tswr}<1$) or positive ($r>1$ and $FF_{m}^{tswr}>1$)
or mixed feedback process ( $r<1$ and $FF_{m}^{tswr}>1$; $r>1$
and $FF_{m}^{tswr}<1$). 

It has been observed that the noise or fluctuations in mRNA/protein
level can be detrimental or beneficial for cellular activities \cite{key-18,key-21,key-39E,key-44,key-46}.
Noise in gene expression also plays an important role in cellular
behaviour and disease control \cite{key-39E,key-44,key-45,key-47}.
Some study shows that the appropriate functioning of the cellular
system requires a specific average level of mRNA and protein \cite{key-39B,key-39C,key-39D}.
Recent work shows that reinitiation of transcription has the capability
to reduce the Fano factor below unity i.e., to the sub-Poissonian
regime in the two-state process\cite{key-39}. We now observe that
transcriptional reinitiation has an important role not only in controlling
the Fano factor but also the average at the mRNA level over a wide
region of parameter space. In the two-state gene expression process,
the cellular system can regulate its mean and Fano factor by controlling
the rate constants responsible for random transitions between the
gene states. In that process, the Fano factor can be reduced up to
unity for large $k_{a}$ and small $k_{d}$. But, in presence of reinitiation
of gene transcription, the cellular system can decrease the Fano factor
below unity at a lower value of $k_{a}$. At the same time, the average
mRNA level compared to the two-state process can be increased. This
is very much important to control diseases like haploinsufficiency
\cite{key-39B,key-39C,key-39D,key-39E}. The noise in gene expression
can also be the survival strategy for cells in adverse environmental
conditions \cite{key-44,key-45,key-47}. The reinitiation of gene
expression can also be helpful in such situations by selecting the
higher Fano factor and appropriate mean mRNA and protein levels. It
has been observed that the protein from essential genes has less variability
compared to the non-essential genes \cite{key-47}. The cellular system
adopted many strategies to control the noise in mRNA and protein levels
among which higher transcription rate and lower translation rate is
an important one \cite{key-47,key-48}. The reinitiation of transcription
can also be a crucial step to keep low noise levels at mRNA and protein
levels. However, this noise minimization is found to be energetically
expensive. It has been shown that a specific average and fluctuations
level of mRNA/protein is subject to the energy consumption of cells
\cite{key-49}. To maintain a higher average level and lower fluctuations
requires higher energy consumption \cite{key-49,key-50}. Thus the
low noise and higher average levels are expected to be advantageous
only when the benefit of reducing noise in a particular gene's expression
outweighs this cost. The reinitiation of transcription step in the
gene regulatory network can help the cellular system to choose the
rate constants in such a way so that the energy consumption and mean
and noise in mRNA/protein levels can be optimized. 

\part*{Appendix}

\section*{A. Constitutive gene expression with pre-initiation and initiation
complexes}

\begin{figure}[H]
\begin{centering}
\includegraphics[width=5cm,height=3cm]{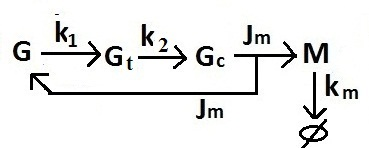}
\par\end{centering}
\caption{\label{fig:gamma dist CWR}Reaction scheme with the rate constants
for constitutive gene expression with pre-initiation and initiation
complexes. $G$ is the open active state and $k_{1}$ is the rate
constant for the open active to the pre-initiation complex formation.
$k_{2}$ is the rate constant for the pre-initiation to the initiation
complex formation. $J_{m}$ is the transcription rate constant and
$k_{m}$ is the mRNA degradation rate constant. }
\end{figure}

The biochemical reactions for constitutive gene expression with pre-initiation
and initiation complexes are shown in figure (\ref{fig:gamma dist CWR}).
The RNAP-II molecule binds the active gene and forms a pre-initiation
complex $G_{t}$. Then further modifications in that produce the initiation
complex from which mRNA synthesis takes place \cite{key-35}. 

Let, there are $l$ copy number of a particular gene exist in the
cell. The Master equation describing the rate of change of probability
$P(n_{1},n_{2},n_{3},t)$ with $n_{1}$ number of mRNAs and $n_{2}$
number of genes in the pre-initiation state ($G_{t}$) and $n_{3}$
number of genes in the initiation complex ($G_{c}$) is given by

\begin{equation}
\begin{array}{ccc}
\frac{\partial p(n_{1},n_{2},n_{3},t)}{\partial t} & = & k_{1}[(l-n_{1}-n_{2}+1)p(n_{1}-1,n_{2},n_{3},t)-(l-n_{1}-n_{2})p(n_{1},n_{2},n_{3,}t)]\\
 &  & +k_{2}[(n_{1}+1)p(n_{1}+1,n_{2}-1,n_{3},t)-n_{1}p(n_{1},n_{2},n_{3,}t)]\\
 &  & +J_{m}[(n_{2}+1)p(n_{1},n_{2}+1,n_{3}-1,t)-n_{2}p(n_{1},n_{2},n_{3},t)]\\
 &  & +k_{m}[(n_{3}+1)p(n_{1},n_{2},n_{3}+1,t)-n_{3}p(n_{1},n_{2},n_{3,}t)]\\
\\
\end{array}\label{eq:A1}
\end{equation}

If the reinitiation process happens in two states (as shown in figure
(\ref{fig:gamma dist CWR})) then the expression of mean mRNA and
Fano factor at mRNA level are given by

\begin{equation}
<m_{r}^{cwrts}>=\frac{k_{1}k_{2}}{J_{m}k_{1}+J_{m}k_{2}+k_{1}k_{2}}\frac{J_{m}}{k_{m}};\label{eq:14.0}
\end{equation}

\begin{equation}
FF_{m}^{cwrts}=1-\frac{J_{m}k_{1}k_{2}(J_{m}+k_{1}+k_{2}+k_{m})}{(J_{m}k_{1}+J_{m}k_{2}+k_{1}k_{2})\{J_{m}(k_{1}+k_{2}+k_{m})+(k_{1}+k_{m})(k_{2}+k_{m})\}}\label{eq:15.0}
\end{equation}

\begin{figure}[H]
\begin{centering}
\includegraphics[width=5cm,height=3cm]{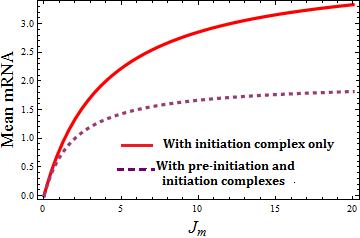}
$\:\:$\includegraphics[width=5cm,height=3cm]{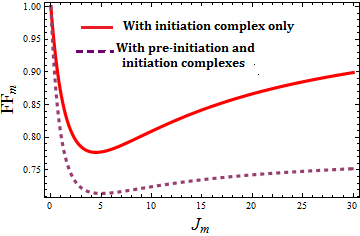}
\par\end{centering}
\caption{\label{fig:CTT-M-FFm}Variation of mean mRNA and Fano factor with
$J_{m}$ for the rate constants $k_{1}=k_{2}=4$ and $k_{m}=1.$}
\end{figure}

Figure (\ref{fig:CTT-M-FFm}) shows that consideration of the pre-initiation
complex in the transcription initiation process results the decrease
in mean and Fano factor further.

\subsection*{B. Two-state gene expression with pre-initiation and initiation complexes}

\begin{figure}[H]
\begin{centering}
\includegraphics[width=9cm,height=3cm]{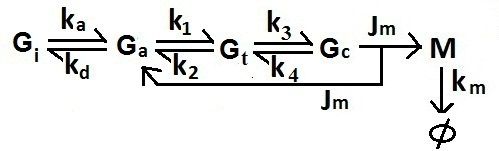}
\par\end{centering}
\caption{\label{fig:Two-state-ga-with-two-transient-states}Reaction scheme
for the two-state gene expression with pre-initiation and initiation
complexes. $G_{i}$ ($G_{a}$) is the inactive (active) state. $k_{a}$
($k_{d}$) is the activation (deactivation) rate constant. $k_{1}$($k_{3}$)
is the rate constant of pre-initiation (initiation) complex formation
and $k_{2}$ ($k_{4}$) is the rate constant of dissociation of RNAP-II
from pre-initiation (initiation) complex. $J_{m}$ is the transcription
rate constant and $k_{m}$ is the mRNA degradation rate constant. }
\end{figure}

The biochemical reactions for the two-state gene activation process
with pre-initiation and initiation steps are shown in figure (\ref{fig:Two-state-ga-with-two-transient-states}).
The gene state $G_{t}$ is the pre-initiation complex and $G_{c}$
is the initiation complex. From the initiation complex, the RNAP-II
starts transcription for mRNA synthesis and the gene turns into an
open active state.

Let $p(n_{1},n_{2},n_{3},n_{4},t)$ be the probability that at time
$t$, there are $n_{4}$ number of mRNAs with $n_{1}$ number of genes
in the active state ($G_{a}$) , $n_{2}$ number of genes in the pre-initiation
state ($G_{t}$) and $n_{3}$ number of genes in the transcription
initiation complex ($G_{c}$). The number of gene in the inactive
state ($G_{i}$) are $(l-n_{1}-n_{2}-n_{3})$ with $l$ being the
copy number of a particular gene. The time evaluation of the probability
is given by the Master equation 

\begin{equation}
\begin{array}{ccc}
\frac{\partial p(n_{1},n_{2},n_{3},n_{4},t)}{\partial t} & = & k_{a}[\{l-(n_{1}-1+n_{2}+n_{3})\}p(n_{1}-1,n_{2},n_{3},n_{4},t)\\
 &  & -\{l-(n_{1}+n_{2}+n_{3})\}p(n_{1},n_{2},n_{3},n_{4},t)]\\
 &  & +k_{d}[(n_{1}+1)p(n_{1}+1,n_{2},n_{3},n_{4},t)-n_{1}p(n_{1},n_{2},n_{3},n_{4},t)]\\
 &  & +k_{1}[(n_{1}+1)p(n_{1}+1,n_{2}-1,n_{3},n_{4},t)-n_{1}p(n_{1},n_{2},n_{3},n_{4},t)]\\
 &  & +k_{2}[(n_{2}+1)p(n_{1}-1,n_{2}+1,n_{3},n_{4},t)-n_{2}p(n_{1},n_{2},n_{3},n_{4},t)]\\
 &  & +k_{3}[(n_{2}+1)p(n_{1},n_{2}+1,n_{3}-1,n_{4},t)-n_{2}p(n_{1},n_{2},n_{3},n_{4},t)]\\
 &  & +k_{4}[(n_{3}+1)p(n_{1},n_{2}-1,n_{3}+1,n_{4},t)-n_{3}p(n_{1},n_{2},n_{3},n_{4},t)]\\
 &  & +J_{m}[(n_{3}+1)p(n_{1}-1,n_{2},n_{3}+1,n_{4}-1,t)-n_{3}p(n_{1},n_{2},n_{3},n_{4},t)]\\
 &  & +k_{m}[(n_{4}+1)p(n_{1},n_{2},n_{3},n_{4}+1,t)-n_{4}p(n_{1},n_{2},n_{3},n_{4},t)]\\
\\
\\
\end{array}\label{eq:25}
\end{equation}

The expressions for mean mRNA and Fano factor for mRNA ( $FF_{m}$)
are given by

\begin{equation}
<m^{tswrth}>=\frac{k_{a}}{k_{m}}\frac{b_{0}}{(b_{1}+b_{3})}\label{eq:26}
\end{equation}

\begin{equation}
FF_{m}^{tswrth}=1-<m^{tswrth}>+\frac{(k_{a}+k_{m})b_{0}}{[k_{m}\{b_{7}+J_{m}(k_{d}(b_{5}-k_{1})+(k_{a}+k_{m})b_{5})+k_{a}(k_{2}(k_{4}+k_{m})+b_{6}(k_{1}+k_{m}))\}]}\label{eq:27}
\end{equation}

where $b_{0}=J_{m}k_{1}k_{3},$ $b_{1}=J_{m}(k_{a}k_{1}+k_{a}k_{2}+k_{d}k_{2}+k_{a}k_{3}+k_{d}k_{3})$,
$b_{2}=k_{m}(k_{d}k_{2}+k_{d}k_{3}+k_{1}k_{3}+k_{d}k_{4}+k_{1}k_{4}+k_{2}k_{4})$,
$b_{3}=k_{a}(k_{1}k_{3}+k_{1}k_{4}+k_{2}k_{4})+k_{d}k_{2}k_{4}$,
$b_{4}=k_{m}^{2}(k_{d}+k_{1}+k_{2}+k_{3}+k_{4}+k_{m})$, $b_{5}=(k_{1}+k_{2}+k_{3}+k_{m})$,
$b_{6}=(k_{3}+k_{4}+k_{m})$, $b_{7}=k_{d}k_{2}k_{4}+b_{2}+b_{4}$.

\begin{figure}[H]
\begin{centering}
\includegraphics[width=5cm,height=3cm]{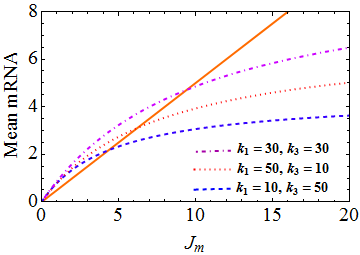} \includegraphics[width=5cm,height=3cm]{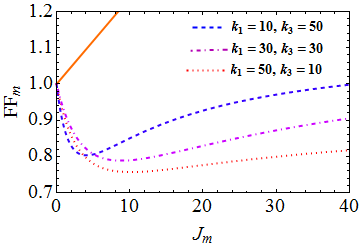}
\par\end{centering}
\caption{\label{fig:mean and FFm TTS}Plot of mean mRNA and Fano factor corresponding
to the figure (\ref{fig:Two-state-ga-with-two-transient-states})
with $J_{m}$ for the rate constants $k_{a}=k_{d}=10,$ $k_{2}=k_{4}=1,$
$k_{m}=1$ and three different sets of $k_{1}$ and $k_{3}$. The
solid line corresponds to the two-state process without reinitiation
(Fig. \ref{fig:Two-state}(a)). }
\end{figure}

Figure \ref{fig:mean and FFm TTS} shows that the mean and the Fano
factor can be controlled in better ways by controlling the rate constants
$k_{1}$ and $k_{3}$.

\end{document}